\documentclass[aps,pra,showpacs,twocolumn,superscriptaddress]{revtex4-1}

\usepackage{bm}
\usepackage{amsmath}
\usepackage{graphicx}
\usepackage{epstopdf}
\usepackage{subfig}
\usepackage{natbib}
\usepackage{epsfig}
\usepackage{amsfonts}
\usepackage{mathrsfs}
\usepackage[toc,page,title,titletoc,header]{appendix}
\usepackage[colorlinks,linkcolor=blue,citecolor=blue,anchorcolor=blue]{hyperref}
\usepackage{dsfont,amsthm,amsbsy}

\usepackage{tikz}
\usepackage{ulem}
\usetikzlibrary{positioning,calc,decorations.pathreplacing}
\usetikzlibrary{arrows,chains,scopes}
\usetikzlibrary{matrix,calc,shapes}
\tikzset{
	ten/.style = {rectangle, draw=black},
	mat/.style = {ten, fill=blue!10, minimum size=1.2em},
	smat/.style={mat, minimum width=2.8em, node font=\small}, 
	fmat/.style ={ten, fill=blue!10, minimum width = 4.7em},
	node distance = 0.5em,
	uni/.style={circle, draw=black, minimum size=3em, fill=yellow!10},
	bguni/.style={uni, minimum size=3.9em},
	node font=\tiny
}

\DeclareMathOperator{\Tr}{Tr}
\newcommand{\dd}{\mathrm{d}}
\newcommand{\ii}{\mathrm{i}}
\newcommand{\e}{\mathrm{e}}

\begin{document}
\title{Scalable Quantum Tomography with Fidelity Estimation}
\author{Jun Wang}
\affiliation{School of Physics, Peking University, Beijing 100871, China}
\author{Zhao-Yu Han}
\affiliation{Department of Physics, Stanford University,
Stanford, California 94305, USA}
\author{Song-Bo Wang}
\affiliation{School of Physics, Peking University, Beijing 100871, China}
\author{Zeyang Li}
\affiliation{Department of Physics, MIT-Harvard Center for Ultracold Atoms and Research Laboratory of Electronics, Massachusetts Institute of Technology, Cambridge, Massachusetts 02139, USA}

\author{Liang-Zhu Mu}
\email{muliangzhu@pku.edu.cn}
\affiliation{School of Physics, Peking University, Beijing 100871, China}
\author{Heng Fan}
\email{hfan@iphy.ac.cn}
\affiliation{Institute of Physics, Chinese Academy of Sciences, Beijing 100190, China}
\affiliation{CAS Central for Excellence in Topological Quantum Computation, University of Chinese Academy of Sciences, Beijing 100190, China}
\author{Lei Wang}
\email{wanglei@iphy.ac.cn}
\affiliation{Institute of Physics, Chinese Academy of Sciences, Beijing 100190, China}
\affiliation{CAS Central for Excellence in Topological Quantum Computation, University of Chinese Academy of Sciences, Beijing 100190, China}

\begin{abstract}
We propose a quantum tomography scheme for pure qudit systems which adopts random base measurements and generative learning methods, along with a built-in fidelity estimation approach to assess the reliability of the tomographic states. We prove the validity of the scheme theoretically, and we perform numerically simulated experiments on several target states including three typical 
quantum information states and randomly initiated states, demonstrating its efficiency and robustness. The number of replicas required by a certain convergence criterion grows in the manner of low-degree polynomial when the system scales, thus the scheme achieves high scalability that is crucial for practical quantum state tomography.
\end{abstract}

\maketitle



\section{Introduction}

With the fast developing techniques of fabricating
quantum devices, we can now manipulate a growing number of entangled qudits. Medium sized quantum devices (10$\sim$100 qubits) have been implemented in the platforms of superconducting circuits, trapped ions and ultra-cold atoms~\cite{ti,Kelly,Monz,HHWang1,HHWang2,lukin,monroe}
. 
Quantum state tomography (QST), which aims at reconstructing an unknown quantum state from suitable measurements on replicas of the state, is a gold standard for verifying and benchmarking the merits of the implementations. 
In particular, QST is necessary for proving the completeness of information that could be provided by all practical 
operations and measurements on a quantum processor. 

Early studies of QST focused on mixed states and found that it requires the information provided by projective measurements on a minimal set of $\mathcal{O}(d)$ mutually unbiased bases~\cite{mub1,mub2,mub3} or by $\mathcal{O}(d^2)$ expectations of positive-operator-valued measures (POVMs)~\cite{mx1,mx2,mx3,mx4}. This soon becomes impractical as Hilbert space dimension $d$ grows exponentially with the number of constituents (e.g. particles). For pure states, it was recently proved that in terms of information the adequate number of POVMs can be drastically reduced to $\mathcal{O}(d)$~\cite{pu1,pu2,pu3} and that of measurement bases can be reduced to four~\cite{five0,five1,five2}. However, it is still experimentally intractable to realize these delicately designed non-local measurements and to acquire corresponding converged probability distributions, since the size of the sample space $d$ is exponentially large~\cite{dis1}. 

After a long history of developing its mathematical ground, we are now at a stage to consider the pragmatic aspects of QST. Specifically, for example, the time efficiency, which is largely determined by the cost of preparing multiple copies of the state and measurements, is crucial for large-scale many-body states tomography. There have been several efforts towards scalable QST schemes~\cite{eff2010,eff13,eff2016,qi2017adaptive,1367-2630-19-11-113017,exp,Herald,efffield,effprl}, mostly by the mean of exploiting the property of short-range entanglement in a matrix product state (MPS)~\cite{mps}. Some colleagues~\cite{NN,CS,kalev2015quantum,pac1,pac2,pac3,Rocchetto2017Learning, Flammia2012, AdaptBayes-PRA,selfguide-PRL113, kyrillidis2018provable} 
applied theories and techniques such as compressed sensing~\cite{CS11,CS22} from the thriving field of machine learning  to this classical problem. We note that there are indeed similarities between QST and unsupervised machine learning tasks such as density estimation~\cite{deeplearningbook}. In both tasks, one aims at modeling high-dimensional probability functions from observed data. While QST is more complicated because probability distributions under different bases are inherently related and modeled simultaneously. 

Besides the efficiency issue, another important practical concern called fidelity estimation~\cite{fe,fe2,Pan} is drawing attention, that is, how to assess the proximity between the tomographic state and the target state in laboratory?

Our work tries to resolve both problems as concentrating on reconstructing pure states in qudit systems through local projective measurements. Viewing the whole quantum state as an integrated generative model and QST as an unsupervised learning task, we make innovations in its key components, measurement and reconstruction. We design a compressed-sensing-inspired approach to acquire data via single-shot measurements on random bases, which avoids full statistics on any specific set of physical observables. Adapting MPS and the associated optimization algorithm proposed by our previous work~\cite{paper}  as the model and learning approach respectively, we devise a cost function based on the averaged negative log-likelihood with an entanglement entropy penalty.
All these efforts help the scheme achieve high efficiency in our computer-simulated experiments for several typical target states that have compact MPS description. Moreover, our scheme allows a fidelity estimation approach that requires no overhead in measurements nor {\it priori} knowledge in the target state.

In this paper, we will first present the workflow of the scheme and argue its validity in Sec.~\ref{s2}, then demonstrate its efficiency and robustness in Sec.~\ref{s3} by computer-simulated experiments.


\section{Procedure}\label{s2}
\begin{figure}[t]
\centering
\includegraphics[width = \linewidth]{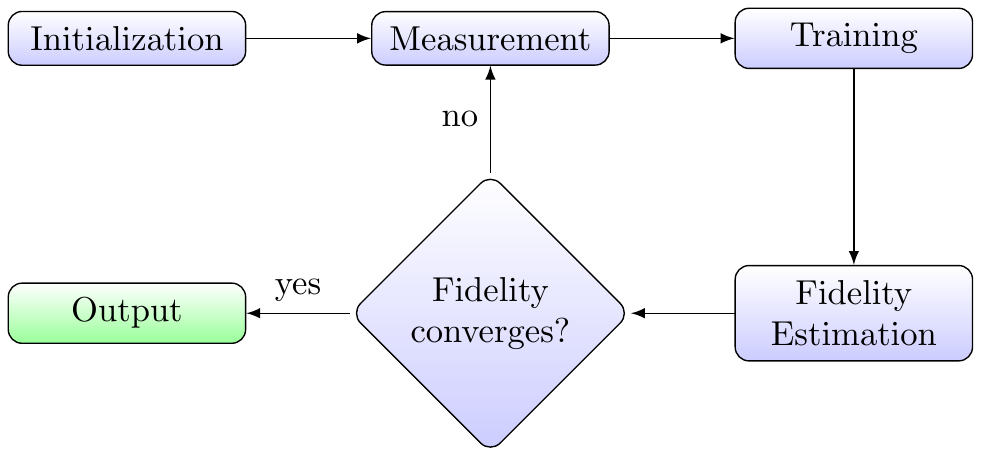}
\caption{A flow chart of the QST scheme.}
\label{mainprocess}
\end{figure}

The proposed QST scheme includes iterative projective measurements on the target state and training of the model, until the stop criterion concerning the fidelity estimation is met, as shown in the flow chart in Fig.~\ref{mainprocess}. The target and the tomographic states are denoted by their density operators $\sigma,\tilde\rho$ respectively. We emphasize that the theoretical arguments in this section are not confined to pure states, although the following experiments are; the general validity of the scheme does not depend on the property of the target state, though the efficiency may.



\subsection{Measurement}\label{sec:measure} 

The bases on which one performs projective measurements are required to comprise an informationally complete set of bases in order to avoid candidate states that are indistinguishable based on the measurement outcomes. In practice, one should also consider the complexity of implementing these measurements. Local measurements are often preferred because of the simplicity of single qudit gates. We note that arbitrary local spin measurements could be expediently realized, e.g. in cold atoms experimental platforms \cite{Keesling2018Lukin}. Here we use the spin-$S$ picture to demonstrate our sets of bases on a qudit system, where the qudits have dimension $q=2S+1$ and $S$ is a half-integer. The base denoted by $\mathcal B (\{\bm n_1,\cdots,\bm n_N\})$, sometimes $\mathcal B (\{\bm n\})$ for short, refers to the eigenbase corresponding to the product of local spin operators 
$\bm n_1\cdot \bm s_1 \otimes \bm n_2\cdot \bm s_2 \otimes \cdots\otimes \bm n_N\cdot \bm s_N$, where $\bm n_i$ is the unit vector indicating the spin measurement direction on site-$i$, and $\bm s_i = (s_i^x,s_i^y,s_i^z)$ are the single site spin operators.
 
We propose to perform measurements on {\it random bases}. This is an analogy of compressed sensing, which recovers a sparse signal by utilizing observations from random perspectives. Each time in demand of a measuring outcome accumulation, one can randomly sample a measuring base $\mathcal B ( \{\bm n\})$, from a certain probability distribution $f(\{\bm n\})$ on $S_2^{\otimes N}$. 

In each projective measurement on base $\mathcal B (\{\bm n\})$, one obtains an outcome $\{m\} = (m_1,m_2,\cdots ,m_N), |m_j|\leq S$ and the repilica state collapses to the corresponding product state $|\{\bm n,m\}\rangle = \otimes_{j=1}^N |\bm n_j,m_j\rangle$. Hence with a certain base sampling distribution, a generic state $\rho$ is naturally a generative model whose sample space comprises all this kind of product states. Note that $|\bm -\bm n_j,-m_j\rangle=|\bm n_j,m_j\rangle$, the probability density is
\begin{equation}\label{eq:mainP}
    \mathbb{P}[\rho](|\{\bm n, m\}\rangle) = \sum_{c_j=\pm 1} f(\{c_j\bm n_j\}_{j=1}^N)\langle \{\bm n, m\}|\rho|\{\bm n, m\}\rangle .
\end{equation}
From this perspective, our scheme is a generative modeling process based on only the outcome product states from the target $\sigma$.
It is proved in Appendix \ref{val} that information consist in the probability density is complete, for the Kullback-Leibler (KL) divergence $D_{\mathrm {KL}}(\mathbb P[\sigma]||\mathbb P[\tilde\rho])$ bounds the matrix norm $||\tilde\rho-\sigma||$ as long as $f$ is not ill-shaped, which means $D_{\mathrm {KL}}(\mathbb P[\sigma]||\mathbb P[\tilde\rho])$ specifies the target as its unique minimum.

Though some schemes have mentioned random base measurements\cite{PhysRevA.93.052105} for QST, ours is radically different in terms of whether a statistical quantity under a certain base is necessary. Since our scheme do not deliberately repeat measurements on any single base, it is possible to avoid the unbearable efforts introduced by complete statistics over exponentially large number of possible outcomes. 

We mention here that in qubit systems, a prescribed set of $\mathcal{O}(N)$ fixed bases corresponding to local measurements also suffices the reconstruction, though it may not achieve high efficiency. See Appendix~\ref{loc} for details.

After accumulating a batch of measurement outcomes, we append them into dataset $\mathcal{V}$, whose size $|\mathcal{V}|$ is also the number of replicas of the target,
and subsequently train the model with the updated $\mathcal{V}$. 

\subsection{Training \label{sec:train}} 

The model for pure states in our scheme is the wave functions under $\mathcal B (\{\bm z_1,\bm z_2, ..., \bm z_N\})$ represented with an MPS. Under another base $\mathcal B (\{\bm n\})$ the wave function is straightforwardly obtained by performing local unitary transformations, schematically:
\begin{equation}\label{PsiMPS}
\tilde\Psi(\{m\}; \{\bm n\}) = 
\begin{tikzpicture}
[baseline=(U1.base)]
{[start chain]
\node [smat,on chain] (A1) at (0,0){$A^{(1)}$};
\node [uni, below=0.3em of A1] (U1) {$U_1$};
\node (r1) [below=0.3em of U1]{$m_1$};
\node [smat,on chain] (A2){$A^{(2)}$};
\node [uni, below=0.3em of A2] (U2) {$U_2$};
\node (r2) [below=0.3em of U2]{$m_2$};
\node [on chain, xshift=-0.3em] (cdt1) {$\cdots$};
\node [smat, on chain,xshift=-0.3em] (AN){$A^{(N)}$};
\node [uni, below=0.3em of AN] (UN) {$U_N$};
\node (rN) [below=0.3em of UN]{$m_N$};
}
\draw (A1)--(U1)--(r1) (A2)--(U2)--(r2) (AN)--(UN)--(rN);
\draw (A1)--(A2)--(cdt1)--(AN);
\end{tikzpicture}.
\end{equation}
where each box denotes a tensor in the MPS, and the circle $U_k$ denotes the single qudit unitary transformation $U(\bm n_k)$ that rotates the direction $\bm n_k$ to $\bm e_z$. For introduction to the graphical notations of tensor networks, we refer to \cite{Schollwock2011, ORUS2014117}. For $\bm n$ in $(\theta,\phi)$ direction in conventional spherical coordinate, $U(\bm n)$ could be realized as 
$
\exp(\mathrm{i}\theta s^y/\hbar) \exp(\mathrm{i}\phi s^z/\hbar)$
.


When the model $\tilde\rho$ minimizes the KL dievergence $D_{\mathrm {KL}}(\mathbb P[\sigma]||\mathbb P[\tilde\rho])$, it is exactly the target state $\sigma$, yet $\sigma$ itself is what we are reconstructing and currently inaccessible. According to Eq.(\ref{eq:mainP}), with both $\sigma, f$ being fixed, it is equivalent for the model to minimize
\begin{equation}
    \mathcal L_N \equiv -\int\dd\{\bm n\}\sum_{\{m\}}\mathbb{P}[\sigma]\ln\langle\{\bm n,m\} |\tilde\rho| \{\bm n,m\}\rangle~.
\end{equation}
Thus with a finite $\mathcal V$, $\mathcal L_N$ is estimated as
\begin{align}
\mathcal L_N = \frac{-1}{|\mathcal{V}|}\sum_{|\{\bm n,m\}\rangle\in \mathcal{V}} 
\ln |\tilde\Psi(\{m\}; \{\bm n\})|^2,\label{eq:LN} 
\end{align}
whose minimum approaches the target $\sigma$ as outcomes accumulate in $\mathcal V$, so we employ it as the cost function for training the MPS. 

$\mathcal L_N$ is indeed highly similar with the negative log-likelihood (NLL) especially when $f$ is uniform, yet generally $|\tilde\Psi(\{m\},\{\bm n\})|^2$ is not the probability density.
Moreover, one can optionally add penalty terms in the cost function $\mathcal L = \mathcal{L}_N+\lambda \mathcal{P}$ as long as he finally converges $\lambda\to 0$. For instance, when training the merged $k^{\mathrm{th}}$ and $(k+1)^{\mathrm{th}}$ tensors, take $\mathcal{P}$ as the second order R\'enyi entropy $-\ln \Tr \rho_{\mathrm{R},k}^2$, where $\rho_{\mathrm{R},k}$ is the reduced density matrix. It renders the MPS preference in low entanglement description of the data in training, and thus make the algorithm converge faster. Encoding information from different bases in the cost function, we may not worry that the MPS overfits the probability distribution on certain bases.

To minimize the cost function, we improve the unsupervised learning approach proposed in our previous work~\cite{paper}. The algorithm is similar with 2-site Density Matrix Renormalization Group \cite{dmrg0} method and adjusts the parameter allocation as well as the parameters of the MPS to the best description of the measured data.

The major difference from the previous work is the gradient calculated for tuning, which results from the complex-value form of quantum wave-functions and the application of unitary transformations. Details about calculating the gradients of $\mathcal L$ are in Appendix \ref{gra}. 






\subsection{Fidelity Estimation}
Estimating the proximity between the tomographic state and the target enables one to determine whether to measure more in practice. Quantum fidelity $\mathcal F\equiv\sqrt{\Tr[\tilde\rho\sigma]}$ and the distance $\mathcal R\equiv||\tilde\rho-\sigma||/2$ are constantly used for this quantification, but we cannot use these definitions for fidelity estimation, since we only have access to the measurement outcomes and the training history instead of the target $\sigma$ itself. 

We assume that similar states' tomography have similar converging behavior when all the scheme parameters are fixed. Thus, when our tomographic state is trained to be a sufficiently good approximation, we use it as a virtual target state $\sigma_\mathrm{vir} = |\tilde\Psi\rangle\langle\tilde\Psi|$ to simulate the tomography scheme in a computer: generate measuring outcomes from $\sigma_\mathrm{vir}$ and train a new model $\tilde\rho_\mathrm{vir}$ in the same way. Since we have access to $\sigma_\mathrm{vir}$, this process could be completely monitored and be used to approximate the real process at the same measurements accumulation stage.

For two pure states, 
$\mathcal{F}=|\langle\psi|\phi\rangle|$ and $\mathcal{R} = \sqrt{\frac{1}{2}(1-\mathcal{F}^2)}$ quantify their proximity. 
In our experimental tests, it is observed that the distance between the tomographic MPS and the target state, $\mathcal{R}_{\text{real}}$, is asymptotically proportional to $|\mathcal{V}|^{-1/2}$, and the distance between two successive tomographic states, $\mathcal{R}_{\text{succ}}$, is asymptotically proportional to $|\mathcal{V}|^{-1}$, at large $|\mathcal{V}|$. This indicates that $\frac{\mathcal{R}^2_{\text{real}}}{\mathcal{R}_{\text{succ}}}$ gradually approaches to a constant $C$ as measurements accumulate. By our assumption that similar states' tomography have close converging processes, we can extract a good approximation of this unknown constant through $\mathcal{R}_{\text{real}}'$ and $\mathcal{R}_{\text{succ}}'$ monitored in the virtual tomography process, so that we can use $\sqrt{C\mathcal{R}_{\text{succ}}}$ to estimate the actual distance $\mathcal{R}_{\text{real}}$ and therefore the fidelity between $\tilde\rho$ and the target $\sigma$. Moreover, one can extrapolate $|\mathcal{V}|$ that is adequate for the fidelity criterion from the asymptote so that the batch size of measurement outcomes could be properly increased to save training efforts. We notice that similar asymptotic behavior of $\mathcal{R}_\mathrm{real}$ has been generally observed in previous QST methods~\cite{PhysRevA.61.042312,PhysRevA.98.012339,PhysRevLett.113.190404}, thus the procedure here may also be adapted for them.

To meet the desired accuracy, we may have to iterate the measurement (Sec.~\ref{sec:measure}) and the training (Sec.~\ref{sec:train}) steps until the convergence criterion of real fidelity is satisfied. When successively trained MPS' as virtual target states give converging estimate of $C$, the confidence in the fidelity estimation also increases.

\section{Experimental Tests}\label{s3}

In this section we test our QST scheme on $N$-qubit systems in computer-simulated experiments. 
We represent the target states with compact MPS'~\cite{je} and employ the direct sampling approach~\cite{paper} to generate independent samples as measurment outcomes.

\subsection{Efficiency}\label{sec:efficiency}
\begin{figure}[ht]
\centering
\includegraphics[width=0.9\linewidth]{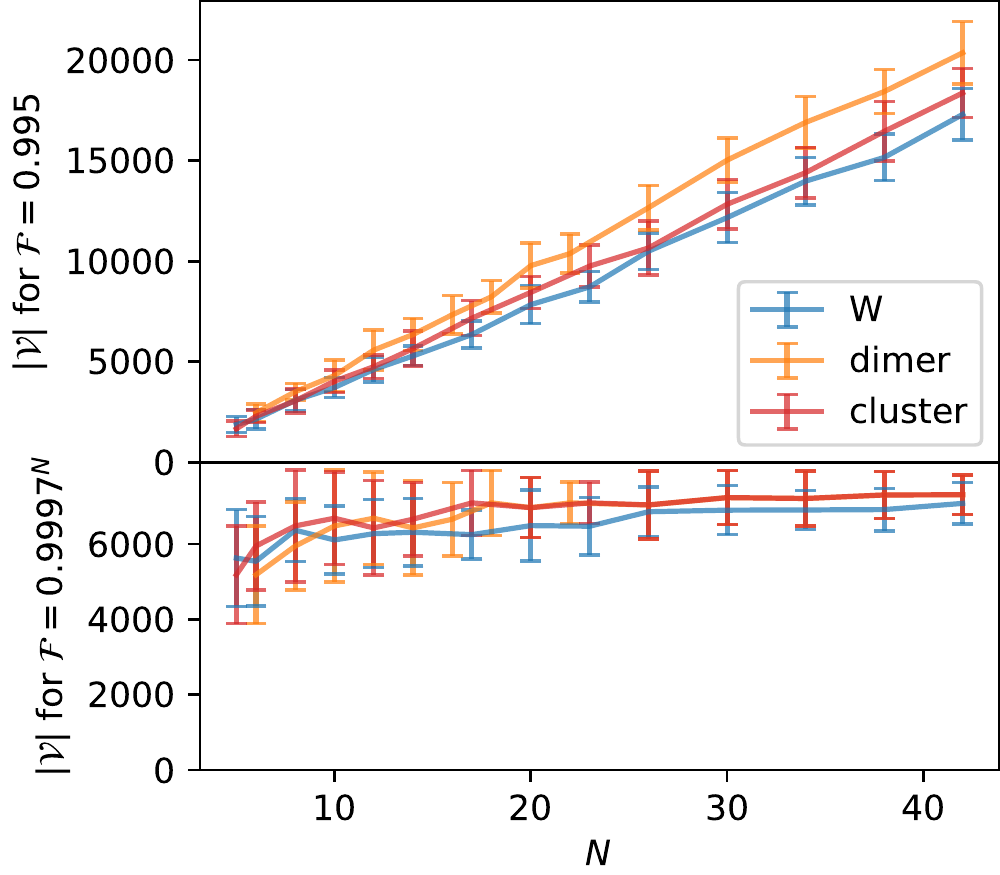}
\caption{Efficiency of the scheme tested on the typical states. The number of single-shot outcomes for a fidelity criterion is recorded when the real fidelity get stably higher than that level. The error bars corresponds to the standard deviation of $24$ cases with different random seeds. 
}
\label{fig:Efficiency}
\end{figure}

Three typical states with compact MPS expression are invoked: W, cluster and dimer states. W states are decorated with local phases 
, $|W\rangle =\sum_{k=1}^N e^{\mathrm{i}k\theta}|0_10_2...1_k...0_N\rangle$, $\theta = 0.1\text{rad}$. 

\begin{figure}[hbtp]
    \centering
    \includegraphics[width=\linewidth]{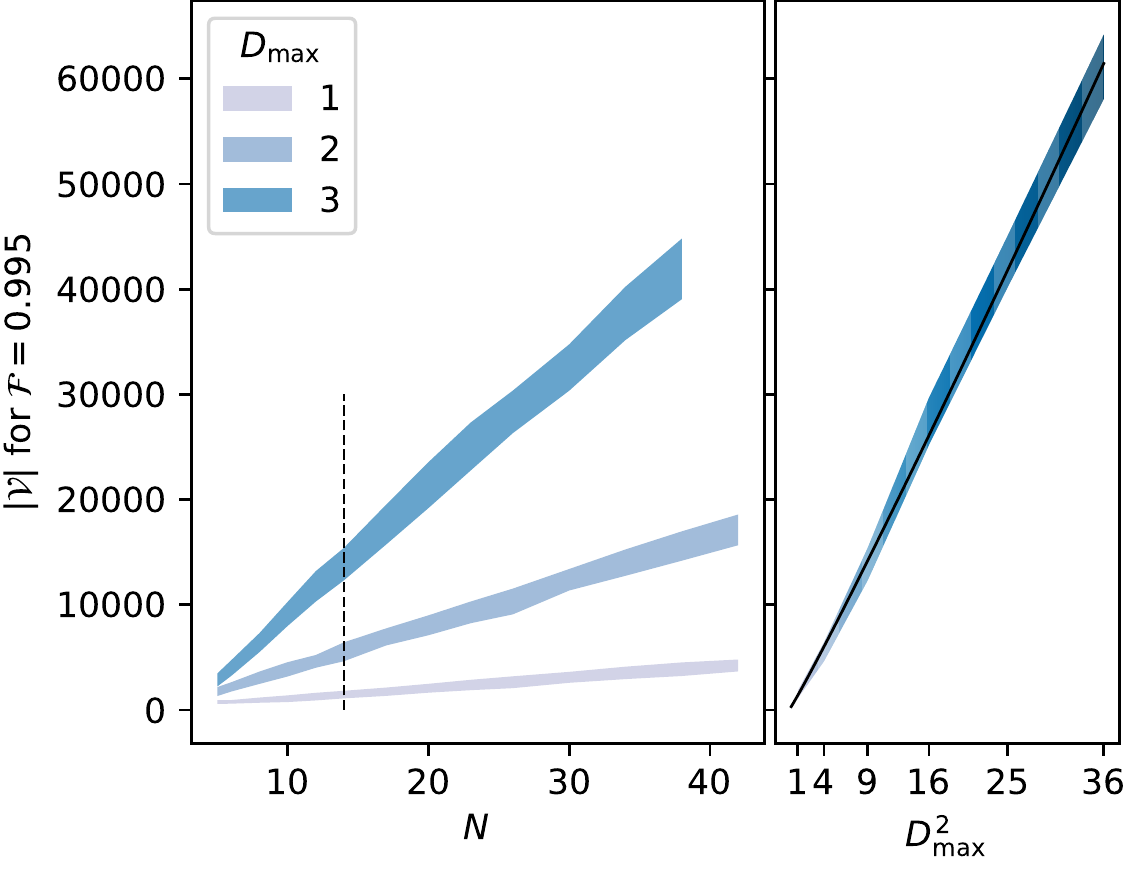}
    \caption{Efficiency of the scheme tested on randomly initiated states. The standard deviation of the number of replicas $|\mathcal{V}|$ over 72 different random targets are covered in the shades. (Left) Linear dependence of $|\mathcal{V}|$ on system size $N$; (Right) Quadratic dependence of $|\mathcal{V}|$ on the maximal bond dimension of the random target states $D_\mathrm{max}$ at $N=14$ (indicated by the dashed line on the left), and the solid line is the fitting $|\mathcal{V}| = \gamma D_\mathrm{max}^\beta$ with the resulting parameters being $\beta=2.1(1),\gamma=7.2(1)$.}
    \label{fig:randinit}
\end{figure}

As shown in Fig.~\ref{fig:Efficiency}, the number of replicas $|\mathcal{V}|$ suffices the criterion of $\mathcal{F} = 0.995$ linearly scales with $N$, and the test on randomly initiated states show the same linear dependence on $N$ in Fig.~\ref{fig:randinit}. To better illustrate the scalability, we consider fixed ``per-site fidelity'' $\mathcal{F}_\mathrm{ps}\equiv\mathcal{F}^{1/N}$, which could be viewed as a normalized criterion for systems with different sizes. In this case, only nearly constant number of replicas are necessary, which is a general corollary 
of the linear dependence combined with the asymptotic behavior of $\mathcal{R}_\mathrm{real}$.
This remarkable relation 
could also be intuitively interpreted with the fact that MPS with finite bond dimensions imposes \emph{sparse} constrains in their Schmidt space so that the compressed-sensing-inspired approach could optimize the utility of the information from the measurements on random bases. To describe a MPS, we only need $\mathcal{O}(N D^2_\mathrm{max})$ parameters, while a single shot projective measurement could provide about $N$ bits of information. Hence the nearly constant demand of measurements is comprehensible. 

To further confirm our argument, we perform experiments on randomly generated target state with maximal bond dimension $D_\mathrm{max}$. 
As it is shown in Fig.~\ref{fig:randinit}, for fixed $\mathcal{F}=0.995$, we observe quadratic and linear scaling respectively when varying $D_\mathrm{max}$ and $\mathcal{N}$, which are exactly what we expected. We thus confirm the both the general validity and scalable complexity of our scheme. 

\subsection{Fidelity Estimation}
It has been generally observed that in our scheme the model has the asymptotic behavior that $\mathcal R_\mathrm{real}\propto|\mathcal V|^{-1/2}, \mathcal R_\mathrm{succ}\propto|\mathcal V|^{-1}$. 
Fitting the asymptotic history into $\mathcal R=C|\mathcal{V}|^{\alpha}$ for all cases in the experiments mentioned above, we obtain $\alpha_\mathrm{real}=-0.52(7),\alpha_\mathrm{succ}=-1.03(4)$, and no dependence on $N$ or $D_\mathrm{max}$ has been observed. These results solidly support our claim about the asymptotic behavior. 
\begin{figure}[ht]
\centering
\includegraphics[width=\linewidth]{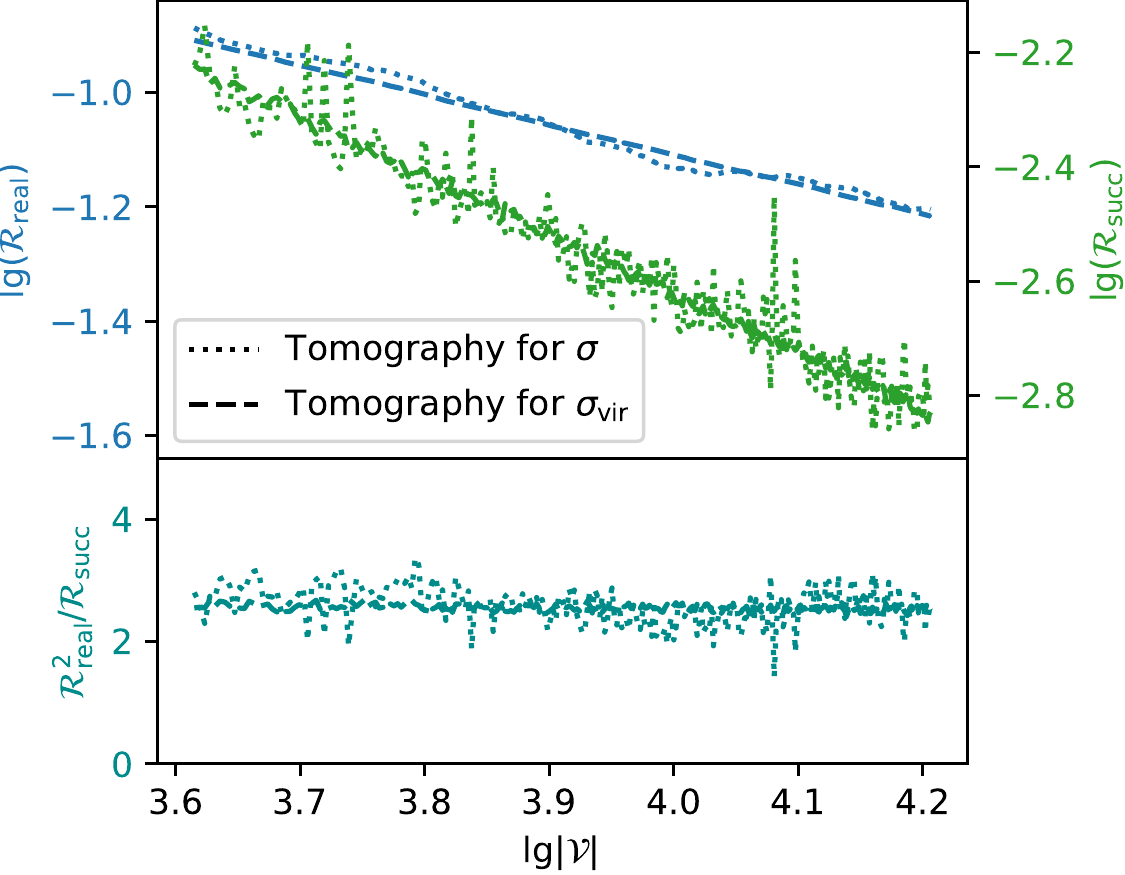}
\caption{Fidelity estimation in the tomography for a $N=30$ W state. A MPS is trained up to $|\mathcal{V}|=10000$ outcomes per our scheme, and $\mathcal{R}_\mathrm{real}, \mathcal{R}_\mathrm{succ}$ and $\mathcal{R}_\mathrm{real}^2/\mathcal{R}_\mathrm{succ}$ in this process are plotted in dotted lines. Then we set it as the virtual target state and simulate our scheme in 24 different random cases. We plot the averaged virtual process in dashed lines.}
\label{fig:CLT}
\end{figure}

Our fidelity estimation approach is tested on those typical states and random states mentioned above. We demonstrate the result on an $N=30$ W state in Fig.~\ref{fig:CLT}, and more results are available in Appendix \ref{sec:FidEst}. Aside from confirming the asymptotic behaviors 
and that $\mathcal{R}_\mathrm{real}^2/\mathcal{R}_\mathrm{succ}$ approaches to a constant, we note that when $\tilde\rho$ is in the vicinity of the target $\sigma$ and plays the role of a virtual target state, it provides a good estimate of the constant $C$ that the actual target state's $\mathcal{R}_\mathrm{real}^2/\mathcal{R}_\mathrm{succ}$ converges to. 
Therefore, with $C$ properly estimated through the virtual process, we are able to assess the real fidelity of our tomographic state to the physical target state.

\subsection{Robustness}
\begin{figure}[ht]
\centering
\includegraphics[width=0.9\linewidth]{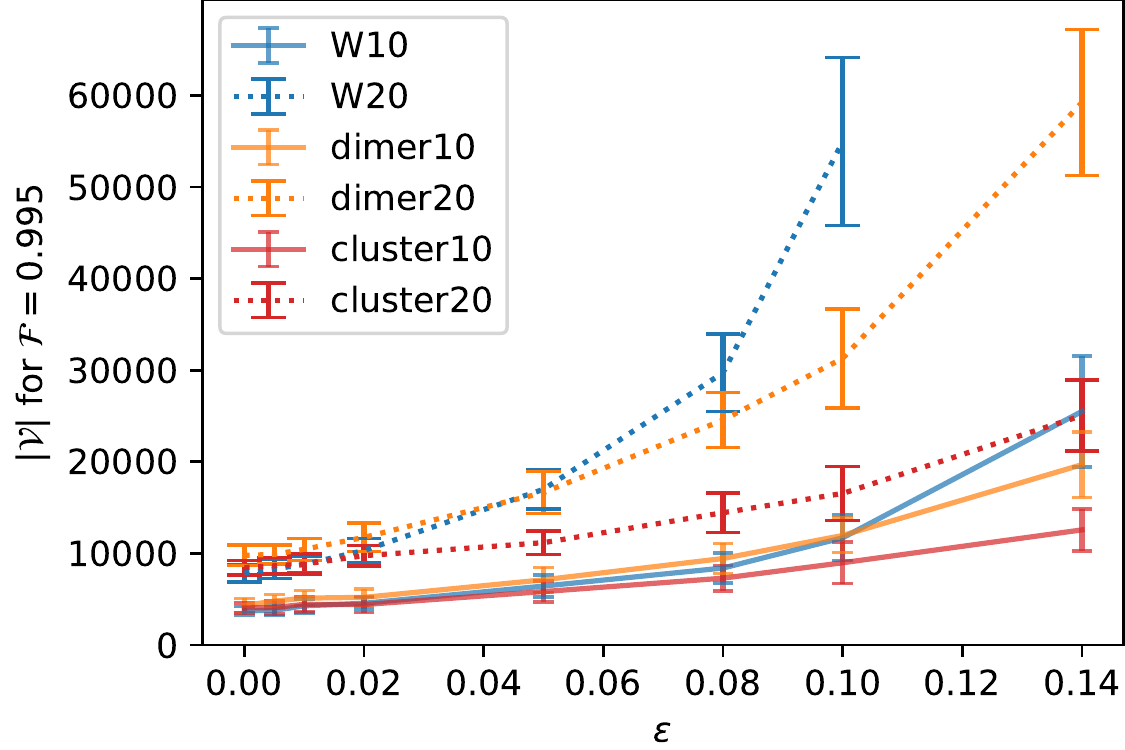}
\caption{Robustness over noises in measurement outcomes. Noise is simulated as: when measured on a base, by a probability of $\epsilon$, the target state generates uniformly random binary strings instead of obeying its probability distribution. The number of replicas necessary for stably $\mathcal{F}\geq 0.995$ are displayed, and the error bars indicate the standard deviation over 24 cases.}
\label{fig:robust}
\end{figure}

Concerning more pragmatic aspects, errors and noises in the implementation of transformations and states are inevitable, and QST is sometimes applied to evaluate the quality of implementation. As it has been shown in the experiments on the typical states and random states, as long as the measurement outcomes obey the probability density, our scheme is acute in the reconstruction of the target state, which means that it can faithfully reflect the systematic deviation of the implementation from what it is supposed to be. Noises in the outcomes that results from the differences between the replicas, however, should be addressed as well. In Fig.~\ref{fig:robust} we demonstrate the robustness of our scheme over the noises. We mix the target with a portion of quantum noise through depolarizing channel, namely the measurement is simulated on
\begin{equation}
    \sigma_\epsilon = (1-\epsilon)\sigma + \frac{\epsilon}{q^N}\mathrm{I},
\end{equation}
where $\mathrm{I}$ is the identity. 
Within $\epsilon<0.05$, the number of outcomes for a certain $\mathcal{F}$ does not significantly rise, yet some states are more sensitive than the others. The asymptotic behaviors are also noticed not to be significantly affected by noises. 

\section{Conclusion and Outlook}
As the size of implemented quantum devices have increased to medium size, the conventional quantum state tomography approaches are on the verge of intractability. We propose a scalable QST scheme for qudit system from an unsupervised learning perspective, which employs projective measurements under random bases, a learning algorithm with MPS model, and a built-in approach to estimate the fidelity of the tomographic state through the process history. The scheme is featured by multiple advantages: the bases setting in our scheme enjoy experimental accessibility of local spin transformations and measurements; the algorithm enables adaptive parameter allocation in the model; the fidelity estimation requires no measurement overhead. Most importantly, the scheme does not acquire statistics on any single base or any set of observables, and thus achieves high efficiency in simulated experiments on the three typical quantum information states and randomly initiated states: the number of replicas needed for prescribed fidelity scales linearly and quadratically when we scales the system size and the maximal entanglement spectra size of the target states, repsectively. We justify the efficacy of the fidelity estimation method and also demonstrated the robustness of the scheme. 

We note that our arguments validating the measurement bases and the optimization function 
are universal for any target quantum states, even for mixed states. We thus expect that our work paves a path to a more general practical tomography scheme by reasonably exploiting other suitable models like multi-scale entanglement renormalization ansatz \cite{PhysRevLett.101.110501, MERA} and neural networks \cite{Carleo602,NN}. Moreover, in a sense that the success of QST marks the full characterization of a quantum device, our work suggests that the more operations we could feasibly choose from, the higher efficiency of information contraction from a quantum resource we would achieve. 

\section*{Acknowledgement}
We thank Xiu-Zhe Luo and Pan Zhang for helpful discussions. This work was supported by Ministry of Science and Technology of China (Grant Nos. 2016YFA0302104 and 2016YFA0300600), 
National Natural Science Foundation of China (Grant Nos. 91536108, 11774406, 11774398). 
J.W. is supported by National Training Program of Innovation for Undergraduates. L.W. and H.F. are supported by Chinese Academy of Sciences (XDPB-0803).  

\section*{Author Contributions}
J.W. and Z.-Y.H. and L.W. developed the procedure. S.-B.W. proved the validity for qubit case, and J.W. generalized it to qudit. J.W., Z.-Y.H. and Z.L. conducted the efficiency experiments. J.W. conducted the virtual tomography and the robustness experiments. L.W., H.F. and L.-Z.M. supervised this work, and all the authors involved in drafting the manuscript. 


\section*{Data Availability}
The data of the simulated experiments mentioned above is available \href{https://github.com/congzlwag/BornMachineTomo}{here} .

\bibliographystyle{apsrev4-1} 
\bibliography{Qpaper}

\onecolumngrid
\appendix

\section{Proof of the Validity of Random Bases} \label{val}
Denoting $\dd\{\bm n\}=(4\pi)^{-N}\dd\bm n_1\ldots\dd\bm n_N$ 
and $\int\dd\{\bm n\}\sum_{m_1,\cdots,m_N}$ as $\int\dd\{\bm n,m\}$ for short, we shall prove that the KL divergence
\begin{equation}
D_{\mathrm {KL}}(\mathbb P[\sigma]||\mathbb P[\tilde\rho]) = \int\dd\{\bm n,m\} \mathbb P[\sigma](\{\bm n,m\}) \ln \frac{\mathbb P[\sigma](\{\bm n,m\})} {\mathbb P[\tilde\rho](\{\bm n,m\})}
\end{equation}
bounds the Schatten norm $||\sigma-\tilde\rho||$.

There are two physical implications. First, because it distinguishes arbitrary two different mixed states, the information it based on is complete for QST; Second, if we know the distribution function \(\mathbb P[\sigma]\) of the target state, then the KL divergence 
is valid for quantifying the proximity between $\tilde\rho$ and $\sigma$, and so is the NLL $-\int\dd\{\bm n,m\} P[\sigma](\{\bm n,m\})\ln P[\tilde\rho](\{\bm n,m\})$ because it only differs from the KL divergence by a constant depending on $\sigma$ and $f$. The distribution \(\mathbb P[\sigma]\) can be attained by an infinite number of measurements, by the law of large number, and the speed of convergence of the distribution is governed by the central limit theorem. 

Pinsker's inequality reads~\cite{Pk}:
\begin{equation}
\sqrt{\frac{1}{2} D_\mathrm{KL} (\mathbb{P}|| \mathbb Q)} \geq \delta (P, Q),
\end{equation}
where \(\delta(\cdot, \cdot)\) is the total variation distance defined by
$\delta(P, Q) = \sup\big\{|P(A) - Q(A)|\big\vert A \text{ is a measurable event}\big\}$,
and $P, Q$ refer to the probability measure given by the probability density functions $\mathbb P, \mathbb Q$, respectively.

Taking an event \(A = \{x\in \text{sample space}~|~\mathbb P(x)\geq \mathbb Q(x)\} \), with $\bar{A}$ being the complementary event, we obtain
\begin{align*}
\int |\mathbb P-\mathbb Q| &= \int_A \mathbb P-\mathbb Q + \int_{\bar  A} \mathbb Q-\mathbb P\nonumber\\ &= \int_A \mathbb P-\int_A \mathbb Q + 1 - \int_A \mathbb Q - 1 + \int_A \mathbb P \nonumber\\&= 2(\int_A \mathbb P-\mathbb Q) =2|P(A)-Q(A)|\leq 2 \delta (P,Q)
\end{align*}

Connecting the two inequalities above, KL divergence bounds the \(L^1\) distance:
\begin{equation}
    \sqrt{2D_\mathrm{KL} (\mathbb{P[\sigma]}|| \mathbb P[\tilde\rho])}\geq
    \int \dd\{\bm n,m\} |\mathbb{P}[\sigma]- \mathbb{P}[\tilde\rho]|
    =\int\dd \{\bm n,m\}F(\{\bm n\}) \big|\langle\{\bm n,m\}|\sigma|\{\bm n,m\}\rangle - \langle\{\bm n,m\}|\tilde\rho|\{\bm n,m\}\rangle\big|,
\end{equation}
where we define $F(\{\bm n\})=\sum_{c_j=\pm}f(\{c_j\bm n_j\}^N_{j=1})$. 

As $\langle\{\bm n,m\}|\rho|\{\bm n,m\}\rangle\geq 0, \sum_{\{m\}}\langle\{\bm n,m\}|\rho|\{\bm n,m\}\rangle=1$, we have $\langle\{\bm n,m\}|\rho|\{\bm n,m\}\rangle\in [0,1]$, thus the $L^1$ distance further bounds the $L^2$ distance between the probabilities
\begin{equation}
\int \dd\{\bm n,m\} |\mathbb{P}[\sigma]- \mathbb{P}[\tilde\rho]| \times 2
\geq \int\dd\{\bm n,m\}F(\{\bm n\}) \big(\langle\{\bm n,m\}|\sigma|\{\bm n,m\}\rangle - \langle\{\bm n,m\}|\tilde\rho|\{\bm n,m\}\rangle\big)^2.
\end{equation}
In order to prove that this \(L^2\) distance bounds the usual matrix distances,
we introduce a product for the linear space of $q^N$ by $q^N$ hermitian matrices ($q=2S+1$):
\begin{equation}\label{eq:inner}
    (\alpha, \beta) \equiv \int\dd\{\bm n,m\}F(\{\bm n\}) \langle\{\bm n,m\} | \alpha |\{\bm n,m\}\rangle \langle\{\bm n,m\} | \beta |\{\bm n,m\}\rangle,
\end{equation}
with which \(\sqrt{(\alpha, \alpha)}\) directly relates to the \(L^2\) distance between probability distributions. Its bi-linearity being obvious, in order to justify it as an inner product we prove its positivity as follow:

As long as $F(\{\bm n\})\geq 0$ is not ill-shaped, $0 = (\alpha,\alpha) $ requires the integrand to be $0$ almost everywhere in $(S_2\times \{S,S-1,\cdots,-S\})^{\otimes N}$, i.e.
\begin{equation}\label{eq:proposition}
    \langle\{\bm n, m\} | \alpha |\{\bm n, m\}\rangle = 0, \forall m\in\{S,S-1,\cdots,-S\}, \text{a.e.} \bm n_j\in S_2 .
\end{equation}
In order to show the conditions in Eq.(\ref{eq:proposition}) restrict $\alpha$ to be $0$, let us look into $N=1$ case first.

\subsection{$N=1$ case}
Denoting $|m\rangle=|\bm z,m\rangle$ for short, with $\bm n$ in $(\theta,\phi)$ direction ($\theta\in[0,\pi],\phi\in[0,2\pi)$) we have
\begin{equation}
|\bm n, m\rangle = \e^{-\ii\phi s^z}\e^{-\ii\theta s^y}\e^{-\ii\varphi s^z} |m\rangle 
= \e^{-\ii m \varphi} \sum_{m'=-S}^S\e^{-\ii m'\phi}|m'\rangle d^{(S)}_{m'm}(\theta)
\end{equation}
where $\varphi$ can be arbitrary, and according to Wigner's formular
\begin{gather}
d_{m'm}^{(S)}(\theta) \equiv \langle m'|\e^{-\ii\theta s^y}|m\rangle = \sum_{k\in\mathbb Z}(-1)^{k+m'-m} W(S,m',m,k) \left(\cos \frac{\theta}{2}\right)^{2S-(2k+m'-m)} \left(\sin \frac{\theta}{2}\right)^{2k+m'-m}\label{eq:Wigner}\\
W(S,m',m,k) \equiv \frac{\sqrt{(S+m')!(S-m')!(S+m)!(S-m)!}}{(S+m-k)!k!(S-k-m')!(k-m+m')!} ,
\end{gather}
where the summation over $k$ takes whatever makes the arguments in the factorials in the denominator are non-negative. Thus $0=\langle \bm n, m |\alpha| \bm n, m\rangle$ gives 
\begin{equation} \label{eq:dad}
0=\sum_{m_1,m_2}\e^{\ii(m_2-m_1)\phi} d_{mm_2}(-\theta)\alpha_{m_2m_1} d_{m_1m}(\theta),~~\text{where } \alpha_{m'm}\equiv \langle m'|\alpha|m\rangle.
\end{equation}
Since $\{\e^{\ii m\phi}|m\in\mathbb Z\}$ is a linear independent set of $\phi\in[0,2\pi)$ functions, in Eq.(\ref{eq:dad}) the coefficient of $\e^{\ii\mu\phi}$ is zero
\begin{equation}
0=\sum_{m_1} d_{m,m_1+\mu}(-\theta) \alpha_{m_1+\mu,m_1} d_{m_1,m}(\theta),
\end{equation}
where $m_1$ runs in the range that $m_1,m_1+\mu\in\{2S,2S-1,\cdots,-2S\}$ in the sum. 

Plugging in Eq.(\ref{eq:Wigner}), $ \forall \mu\in\{2S,2S-1,\cdots,-2S\}, \forall m\in\{S,S-1,\cdots,-S\}$
\begin{align}
0 = \sum_{m_1}\alpha_{m_1+\mu,m_1}\sum_{k_1,k_2} &(-1)^{k_1+m-m_1-\mu}(-1)^{k_2+m_1-m}W(S,m,m_1+\mu,k_1)W(S,m_1,m,k_2) \nonumber\\
& \left(\cos \frac{-\theta}{2}\right)^{2S-(2k_1+m-m_1-\mu)} \left(\sin \frac{-\theta}{2}\right)^{2k_1+m-m_1-\mu} \left(\cos \frac{\theta}{2}\right)^{2S-(2k_2+m_1-m)} \left(\sin \frac{\theta}{2}\right)^{2k_2+m_1-m}\nonumber\\
= \sum_{m_1}\alpha_{m_1+\mu,m_1} \sum_{k_1,k_2} &(-1)^{k_1+k_2+m-m_1} W(S,m,m_1+\mu,k_1)W(S,m_1,m,k_2)\nonumber\\
&\left(\cos \frac{\theta}{2}\right)^{4S-2(k_1+k_2)+\mu} \left(\sin \frac{\theta}{2}\right)^{2(k_1+k_2)-\mu} \label{eq:cossin_base}
\end{align}
$k_1,k_2$ considered in the sum are confined so that in the denominators of
\begin{align}
W(S,m,m_1+\mu,k_1) &= \frac{\sqrt{(S+m)!(S-m)!(S+m_1+\mu)!(S-m_1-\mu)!}}{{\color{purple}(S+m_1+\mu-k_1)!}{\color{blue}k_1!}{\color{purple}(S-k_1-m)!}{\color{blue}(k_1+m-m_1-\mu)!}}\\
W(S,m_1,m,k_2) &= \frac{\sqrt{(S+m_1)!(S-m_1)!(S+m)!(S-m)!}}{{\color{purple}(S+m-k_2)!}{\color{blue}k_2!}{\color{purple}(S-k_2-m_1)!}{\color{blue}(k_2+m_1-m)!}}
\end{align}
there are no negative arguments of the factorials. Thus 
${\color{blue} 2(k_1+k_2)-\mu \geq 0}$ and ${\color{purple} 4S-2(k_1+k_2)+\mu \geq 0}$. Meanwhile, note that $\{\cos^{N-n}x\sin^nx | n=0,1,\cdots ,N\}$ is a linear independent set of $x\in[0,\pi/2]$ functions. Hence $\forall \mu\in\{2S,2S-1,\cdots,-2S\}$ and $\forall k=k_1+k_2\in\{0,1,\cdots,2S\}\cap\{\mu,\mu+1,\cdots,\mu+2S\}$, the coefficient of $\left(\cos \frac{\theta}{2}\right)^{4S-2k+\mu} \left(\sin \frac{\theta}{2}\right)^{2k-\mu}$ in Eq.(\ref{eq:cossin_base}) is zero
\begin{equation}
0 =\sum_{m_1}\alpha_{m_1+\mu,m_1} \sum_{k_1} (-1)^{k+m-m_1} W(S,m,m_1+\mu,k_1)W(S,m_1,m,k-k_1)~.
\end{equation}

Consider conditions required by $m=S$, the $k_1!(-k_1)!$ in the denominator of $W(S,m,m_1+\mu,k_1)$ \textbf{confines} $k_1$ to $0$ in the sum. Similarly, the sum over $m_1$ \textbf{only} takes $m_1=S-k$ because of the $(S-k_2-m_1)!(k_2+m_1-S)!$, thus we confirm 
\begin{equation}
0=\alpha_{S-k+\mu,S-k}(2S)!\left((2S-k+\mu)!(k-\mu)!(2S-k)!k!\right)^{-\frac{1}{2}}
\end{equation}
which suffices $\alpha=0$ as $k,\mu$ runs over whatever they are allowed.

Since all the $q$ by $q$ Hermitian matrix constitute a $q^2$-dimension $\mathbb R$ space, denoted by $\mathcal H$, it is implied that $\{|\bm n, m\rangle\langle\bm n, m| \big | \bm n\in S_2,m\in\{S,S-1,\cdots, -S\}\}$ includes a basis of $\mathcal H$.

\subsection{Back to general $N$ case}
The space of all $q^N$ by $q^N$ Hermitian matrices is $q^{2N}$-dimensional, and it is also $\mathcal H^{\otimes N}$, thus $\{\otimes_{j=1}^N |\bm n_j, m_j\rangle\langle\bm n_j, m_j| \big | \bm n_j\in S_2,m_j\in\{S,S-1,\cdots, -S\}\}$ includes a basis of $\mathcal H^{\otimes N}$. Hence Eq.(\ref{eq:proposition}) suffices $\alpha=0$ .

Thus we confirm that \((\cdot,\cdot)\) is an inner product in the $q^N$ by $q^N$ hermitian matrices space. Since all inner products in the same space are equivalent to each other, $(\alpha,\alpha)=C||\alpha||^2$. In summary,
\begin{align*}
\sqrt{2D_{\mathrm{KL}}(\mathbb P[\sigma]||\mathbb P[\rho])}
\geq& 2\delta(P[\rho], P[\sigma]) \geq \int \dd\{\bm n, m\}\big|\mathbb P[\rho] - \mathbb P[\sigma]\big|\\
\geq& \frac{1}{2}\int\dd\{\bm n,m\}F(\{\bm n\}) \big(\langle\{\bm n,m\}|\sigma-\tilde\rho|\{\bm n,m\}\rangle\big)^2\\
=& \frac{C}{2}||\sigma-\tilde\rho||^2
\end{align*}
\section{An Alternative Way of Choosing Bases in Qubit System} \label{loc}
Due to the limitation of experimental apparatus or the feature of reconstruction algorithm, one may prefer measurements on a prescribed set of {fixed bases} constituting of local measurement along certain directions, for example, the $2N+1$ bases used in~\cite{NN}: 
the first base $\mathcal B(\{\bm z_1,\bm z_2, ..., \bm z_N\})$ determines the amplitudes of the wave-function, and the other $2N$ bases, $\mathcal{B}_{x,y}^k$ rotate one of the $N$ directions from $\bm z$ to either $\bm x$ or $\bm y$, determining the local relative phases of the state. We prove completeness of the information that could be provided by this set of bases in this appendix. 

We first choose the standard base in \(N\)-qubit state space as \(\{|i_1 i_2 ... i_N\rangle\}\), which are eigenstates of \(\sigma_1^z\otimes\sigma_2^z\otimes ...\otimes \sigma_N^z\). In our construction of local operators, it is important to think this set of base as a \(N\)-dimensional cube, and the adjacency relation (quantum mechanically a flip of spin) will be exploited. For the purpose of notational convenience, these basis vectors, or vertices of \(N\)-dimensional cube \(\mathcal C\), are identified with a \(\mathbb{Z}/ 2\mathbb{Z}\) vector space: \(\mathcal{C}_N \cong (\mathbb{Z}/2\mathbb Z)^N\). That is to say, we allow the addition between the labels \(i_1 \cdots i_N\) for the base vectors, with periodic boundary condition applied. For instance, the label for all down spin \(1 1\cdots 1\) added by the label \(1 0 \cdots 0\) will be \(0 1 \cdots 1\), because \(2 = 0 \mod 2\). One can see this addition operation is equivalent to applying a flip operator \(\sigma^x\) on the first spin. If we go back to the \(N-\)dimensional cube, the operation gives a adjacent vertex. For convenience, we define a set of unit vector \(\{e_i\}_{1\leq i\leq N}\) in \(\mathcal C_N\), where \(e_i\) means \(1\) in the \(i\)-th digit, and \(0\) in all others. Suppose the unknown pure state is \(|\psi\rangle = \sum_{i_1 \cdots i_n}c_{i_1\cdots i_n}|i_1\cdots i_n\rangle = \sum_{v \in \mathcal C_n} c_v |v\rangle\). In the following we will allow addition on the state labels, i.e. \(|v+w\rangle\) represents another state in \(N\)-qubit space if \(|v\rangle, |w\rangle\) are also states in the space. For example, \(|11 \cdots 1 + 10 \cdots 0\rangle = |01\cdots 1\rangle\). 

Now we define a set of hermitian operators \(\{T^k_v, \tilde T^k_v\}_{1\leq k\leq N, v\in \mathcal C_N}\) which helps us determine the quantum state.
\[T^k_v = |v\rangle\langle v+e_k| + |v+e_k\rangle \langle v|\]
\[\tilde T^k_v = - i |v\rangle\langle v+e_k| +i |v+e_k\rangle \langle v|\]
Taking the expectation value of these operators under \(|\psi\rangle\) gives
\[\langle \psi | T^k_v | \psi \rangle = c_v^* c_{v+e_k} + \mathrm{c.c.} \]
\[\langle \psi | \tilde{T}^k_v | \psi \rangle = - i c_v^* c_{v+e_k} + \mathrm{c.c} \]
\begin{equation}
\label{connect}
2c_v^*c_{v+e_k} = \langle \psi | T^k_v|\psi\rangle + i\langle \psi | \tilde T^k_v|\psi\rangle 
\end{equation}
Now we write \(T^k_v, \tilde T^k_v\) in terms of rank-\(1\) operators, i.e. density matrices for a pure state.
\begin{align*}
T^k_v =& \frac 12(|v\rangle + |v+e_k\rangle)(\langle v| + \langle v+e_i|) \\&- \frac 12(|v\rangle - |v+e_k\rangle)(\langle v| - \langle v+e_i|) \\ 
\tilde T^k_v =& \frac 12(|v\rangle + i|v+e_k\rangle)(\langle v| -i \langle v+e_i|) \\&- \frac 12(|v\rangle - i|v+e_k\rangle)(\langle v| +i \langle v+e_i|)
\end{align*}
Then the bases \(\mathcal B^k_x = \{|v\rangle + |v+e_k\rangle, |v\rangle - |v+e_k\rangle\}, \mathcal B^k_y = \{|v\rangle +i |v+e_k\rangle, |v\rangle - i|v+e_k\rangle\}\) gives complete information of the Eq.~(\ref{connect}). We notice that \(\mathcal B^k_x\), \(\mathcal B^k_y\) are eigenvectors of the local measuring operator \(\sigma^z_1\otimes\sigma^z_2\otimes...\otimes \sigma^z_{k-1}\otimes\sigma^x_k\otimes\sigma^z_{k+1}\otimes...\otimes\sigma^z_{N}\), and \(\sigma^z_1\otimes\sigma^z_2\otimes...\otimes \sigma^z_{k-1}\otimes\sigma^y_k\otimes\sigma^z_{k+1}\otimes...\otimes\sigma^z_{N}\) respectively. Thus of total we have \(2N+1\) such local measuring operators.
 
In the case of \(c_v, c_{v+e_k} \neq 0\), the preceding equation allows us to compute one from another. Diagrammatically we draw an edge connecting \(v\) and \(v+e_k\). Do this repeatedly for all \(v\), and finally we get a graph.  Alternatively we can also consider the vertices \(v\) with coefficient \(c_v = 0\), and by our procedure, those vertices are exactly those whose edges are missing in the final graph. Then one can for exmaple suppose \(c_{0 0\cdots 0} = 1\), and compute the coefficients of the vertices connected to \(0 0\cdots 0\), by our equation of \(c_v\) and \(c_{v+e_k}\). Maximally extended, the coefficients in the connected component of \(00 \cdots 0\) are determined. We then see if the graph is connected, i.e. the only connected component by the graph itself, the state is determined up to an overall constant. If the graph is not connected, the coefficient in each component is determined up to an overall constant, but the ratio between the overall constants are undetermined. To be more clear, we choose representatives \(r_1, \ldots, r_k\) from each of the connected components. Once we know \(c_{r_i}\), the coefficients of the connected component containing \(r_i\) are known. Then in order to determine the state, what remains to be done is to determine the ratios between \(c_{r_i}\) and \(c_{r_j}\). 

Define \(\Omega\) as the set of quantum state with no zero coefficient \(c_v \neq 0\) for any \(v\). The states in \(\Omega\) can be uniquely determined, according to the previous discussion, and the complement \(A = \mathcal H - \Omega\) is a finite union of \(2^N-1\) dimensional hyperplane \(A_v = \{|\psi\rangle| \langle v|\psi\rangle =0 \}\), thus is of measure zero. Projective measurement of operator \(\sigma^z_1\otimes\sigma^z_2\otimes...\otimes\sigma^z_{N}\) will determine whether the state is in \(\Omega\). Thus we have constructed \(2N+1\) local measurements valid for determination of almost all states.

Since the \( ({\mathbb Z/2 \mathbb Z})^N\) is abelian and has exponent \(2\), the least number of generators for this group is \(N\). Then if we wish to use another set of vectors \(f_j\) instead of \(e_i\) to make the graph connected, we still need \(N\) vectors, and the \(2N+1\) result cannot be improved in this sense.

We now discuss the measure zero set \(A = \cup_{v\in \mathcal C_n}A_v\) a little bit further. We now allow a experimental and computational error of \(\varepsilon\), which means if \(|c_v| < \varepsilon\), we will regard it as \(c_v = 0\). Such vectors form a set \(A^\varepsilon\), and \(A^\varepsilon \subset \bigcup_{v\in \mathcal C_N}A^\varepsilon_v)\), where \(A^\varepsilon_v\) is the tabular neighborhood of \(A_v\) of radius \(\varepsilon\). Suppose after normalizing vectors in \(A, A^\varepsilon, A^\varepsilon_v\) we get \(B, B^\varepsilon, B^\varepsilon_v\) respectively, the probability of finding a vector in the normalized Hilbert space is
\begin{align*}
P(A^\varepsilon) = \frac{\mathrm{Vol} B^\varepsilon}{\alpha(2 \times 2^N)} &\leq \frac{2^N \mathrm{Vol} B^\varepsilon_v}{\alpha(2\times 2^N)}\\&\leq \frac{2^N\alpha(2(2^N-1))\pi \varepsilon^2}{\alpha(2^N)} = \varepsilon^2
\end{align*}
where \(\alpha(k)\) is the volume for \(k\)-dimensional ball. 

In order to reduce the measure of the states we cannot determine, we rotate the whole system by some axis and angle. For example, in our practical choice of the bases, the $\bm{z}$ axis is rotated to former $\bm{x}$ and $\bm{y}$ axises. We denote the state with small components in the new coordinate system as \(N'^\varepsilon\). If the rotation is ``large" in \(SU(2^N)\), we could expect the two events \(A^\varepsilon\) and \(A'^\varepsilon\) are independent, thus the probability is \(P(A^\varepsilon\cap A'^\varepsilon) = \varepsilon^4\). By performing the ``large" rotations repeatedly for \(K\) times, the probability can hopefully be reduced to order \(\varepsilon^{2K}\). This conclusion indicates that even when considering the possible error in the experiment and computation, the expectation of the number of needed bases is still bounded by \(O(N)\).


Furthermore, there are states whose coordinates do not change under any spatial rotations, i.e. total spin \(0\) states. If those states happen to have zero components and the corresponding graphs happen to be disconnected, they can not be uniquely determined by any single set of bases we proposed. One simplest example of such states is the singlet state \(\frac{1}{\sqrt 2}(|10\rangle - |01\rangle)\). The state has total spin 0, thus its components are not changed under rotation. The corresponding graph is \(\mathcal B_2\) with vertices \(00, 11\) removed, and is disconnected. While this can be fixed, by combining the information from measurements under several rotated sets of bases.

We note here that, in experimental tests, we found that sometimes exponentially large amount of measurements are needed for the convergence of probability on these {fixed bases}. For example, the product state $\left(\frac{|0\rangle+|1\rangle}{\sqrt[]{2}}\right)^{\otimes N}$ has at least $2^{N-1}$ non-zero component on every base, so the needed number of measurements grows exponentially in $N$ to converge. Moreover, performing tomography on these fixed bases can be biased for finite dataset since there exist special space directions in the measurements.

\section{Computing the Gradient}\label{gra}
A state of an $N$-qudit system has a general form:
\begin{equation}
\vert \Psi\rangle = \sum_{v_1, v_2,\cdots, v_N } \Psi_{v_1, v_2,\cdots, v_N} \vert v_1 v_2 \cdots v_N\rangle
\end{equation}
With $v_k\in \{S,S-1,\cdots, -S\}$ convention, $\vert v_1 v_2 \cdots v_N\rangle$ could be an eigenstate of the operator $s^{z}_1 \otimes s^{z}_2 \otimes\cdots\otimes  s^{z}_N$. The tensor $\Psi_{v_1, v_2,\cdots, v_N}$ could be decomposed to a matrix product state (MPS):
\begin{equation}
\Psi_{v_1,v_2,\cdots, v_N} = \sum_{i_1,i_2,\cdots, i_{N-1}}A^{(1)v_1}_{i_1}A^{(2)v_2}_{i_1i_2}\cdots A^{(N-1)v_{N-1}}_{i_{N-2}i_{N-1}}A^{(N)v_N}_{i_{N-1}}
\end{equation}
where the summation of $i_{k}$ runs over $1$ to $\mathcal D_k$. Schematically
\begin{align}
\begin{tikzpicture}[baseline=(Psi.south)]
\node[smat,minimum width= 7em] (Psi) at (0,0) {$\Psi$};
\node (v1) [below= of Psi, xshift=-3.5em] {$v_1$};
\node (v2) [right=of v1] {$v_2$};
\node (vN) [below= of Psi, xshift=3.5em] {$v_N$};
\node (cdots) at ($(v2)!0.5!(vN)$) {$\cdots$};
\draw (v1)--(v1|-Psi.south) (v2)--(v2|-Psi.south) (vN)--(vN|-Psi.south);
{[start chain]
\node (eq) [right= of Psi, on chain] {$=$};
\node[smat, on chain] (A1) {$A^{(1)}$};
\node (v11) [below=of A1 ]{$v_1$};
\node[smat] (A2) [on chain] {$A^{(2)}$};
\node (v22) [below=of A2 ]{$v_2$};
\node (cdts2) [on chain, xshift=-0.3em] {$\cdots$};
\node[smat] (AN) [on chain, xshift=-0.3em] {$A^{(N)}$};
\node (vNN) [below=of AN ]{$v_N$};
}
\draw (v11)--(A1) (v22)--(A2) (vNN)--(AN);
\draw (A1)--(A2) (A2)--(cdts2) (cdts2)--(AN) ;
\end{tikzpicture}
\end{align}
With bond dimensions $\mathcal D_k$ permitted to be sufficiently large, this decomposition is precise.

When the algorithm sweep to bond $k$, the cost function is subjected to a penalty proportional to the R\'{e}nyi entropy $S_{2,k} = -\ln \Tr \left(\rho_{\mathrm{R},k}^2\right)$, where $\rho_{\mathrm{R},k}$ is the reduced density matrix of one of the two parts separated by bond $k$.
\begin{align}
\mathcal{L} =& \mathrm{NLL} + \lambda S_{2,k}\\
=& \frac{-1}{|\mathcal{V}|}\sum_{(\bm r, \{\bm n\})\in \mathcal{V}} 
\ln \frac{|\Psi_{\mathrm{MPS}}(\bm{r}; \{\bm n\})|^{2}}{\mathcal{N}}-\lambda \ln\Tr \left(\rho_{\mathrm{R},k}^2\right)
\end{align}
where
\begin{align}
\mathcal{N}  &=  
\begin{tikzpicture}[baseline=(A1d.north)]
{[start chain]
\node [mat,on chain] (A1) at (0,0){};
\node [mat, on chain] (A2) {};
\node [on chain, xshift=-0.3em] (cdt1) {$\cdots$};
\node [mat, on chain,xshift=-0.3em] (AN){};
}
{[start chain]
\node [mat,on chain, below=of A1] (A1d) {};
\node [mat, on chain] (A2d) {};
\node [on chain, xshift=-0.3em] (cdt1d) {$\cdots$};
\node [mat, on chain,xshift=-0.3em] (ANd) {};
}
\draw (A1)--(A2)--(cdt1)--(AN) (A1d)--(A2d)--(cdt1d)--(ANd) (A1)--(A1d) (A2)--(A2d) (AN)--(ANd);
\end{tikzpicture}\\
\Tr \left(\rho_{\mathrm{R},k}^2\right) &= 
\begin{tikzpicture}[baseline=(A1p.north)]
{[start chain]
\node[mat, on chain] (A1) at (0,0) {};
\node[on chain, xshift=-0.3em] (cdt1) {$\cdots$};
\node [mat, on chain,xshift=-0.3em] (Ak){};
\node [mat, on chain] (Ak1){};
\node [on chain, xshift=-0.3em] (cdt2) {$\cdots$};
\node [mat, on chain,xshift=-0.3em] (AN){};
}
{[start chain]
\node [mat,on chain, below=of A1] (A1d) {};
\node [on chain, xshift=-0.3em] (cdt1d) {$\cdots$};
\node [mat, on chain,xshift=-0.3em] (Akd) {};
\node [mat, on chain] (Ak1d){};
\node [on chain, xshift=-0.3em] (cdt2d) {$\cdots$};
\node [mat, on chain,xshift=-0.3em] (ANd) {};
}
{[start chain]
\node[mat, on chain, below=of A1d] (A1p) {};
\node[on chain, xshift=-0.3em] (cdt1p) {$\cdots$};
\node [mat, on chain,xshift=-0.3em] (Akp){};
\node [mat, on chain] (Ak1p){};
\node [on chain, xshift=-0.3em] (cdt2p) {$\cdots$};
\node [mat, on chain,xshift=-0.3em] (ANp){};
}
{[start chain]
\node [mat,on chain, below=of A1p] (A1dp) {};
\node [on chain, xshift=-0.3em] (cdt1dp) {$\cdots$};
\node [mat, on chain,xshift=-0.3em] (Akdp) {};
\node [mat, on chain] (Ak1dp){};
\node [on chain, xshift=-0.3em] (cdt2dp) {$\cdots$};
\node [mat, on chain,xshift=-0.3em] (ANdp) {};
}
\draw (A1)--(cdt1)--(Ak)--node [above=0.7em](bond annot) {bond $k$} (Ak1)--(cdt2)--(AN)
(A1d)--(cdt1d)--(Akd)--(Ak1d)--(cdt2d)--(ANd)
(A1p)--(cdt1p)--(Akp)--(Ak1p)--(cdt2p)--(ANp) 
(A1dp)--(cdt1dp)--(Akdp)--(Ak1dp)--(cdt2dp)--(ANdp)
(Ak1.north)--++(0,0.1em)--++(1em,0)|-($(Ak1d.south)-(0,0.1em)$)-|(Ak1d.south)
(Ak1p.north)--++(0,0.1em)--++(1em,0)|-($(Ak1dp.south)-(0,0.1em)$)-|(Ak1dp.south)
(AN.north)--++(0,0.1em)--++(1em,0)|-($(ANd.south)-(0,0.1em)$)-|(ANd.south)
(ANp.north)--++(0,0.1em)--++(1em,0)|-($(ANdp.south)-(0,0.1em)$)-|(ANdp.south)
(A1d)--(A1p)
(Akd)--(Akp)
(A1.north)--++(0,0.1em)--++(-1em,0)|-($(A1dp.south)+(0,-0.1em)$)-|(A1dp.south)
(Ak.north)--++(0,0.1em)--++(-1em,0)|-($(Akdp.south)+(0,-0.1em)$)-|(Akdp.south);
\draw[->] (bond annot)--++(0,-0.9em);
\end{tikzpicture}
\end{align}
In practice we take $\lambda$ to be an annealing parameter, so that as the MPS becomes well trained, the penalty is fading away. After merging $A^{(k)}$ and $A^{(k+1)}$ as shown in Eq.(\ref{merging}), gradients are calculated as Eq.(\ref{gradient-main}). Notice that $\Psi$ takes complex value, so in order to decrease $\mathcal{L}$, $A^{(k,k+1)}$ should vary in the direction of $-\partial\mathcal{L}/\partial A^{(k,k+1)*}$.

As in \cite{paper}, we adapt the gauge for MPS so that whenever merging two adjacent matrices, the matrices on their left are left-canonical and right-canonical on their right. As long as this gauge is kept, tensors such as $\mathcal{N}, \Tr \left(\rho_{\mathrm{R},k}^2\right)$ that are for the calculation of gradient are greatly simplified as shown in Eqs.(\ref{eq: Norm})-(\ref{eq: Tprime}).

\begin{gather}
\begin{tikzpicture}[baseline=(Ak.south)]
\node[smat] (Ak) at (0,0) {$A^{(k)}$};
\node[smat, right=of Ak] (Akp1) {$A^{(k+1)}$};
\node (vk) [below=of Ak] {$v_k$};
\node (vkp1) [below=of Akp1] {$v_{k+1}$};
\node (ikm1) [left=of Ak] {$i_{k-1}$};
\node (ikp1) [right=of Akp1] {$i_{k+1}$};
\draw (vk)--(Ak) (vkp1)--(Akp1) (ikm1)--(Ak) (ikp1)--(Akp1);
\draw [] (Ak)--(Akp1);
\end{tikzpicture}
=
\begin{tikzpicture}[baseline=(Akkp1.south)]
\node [smat] (Akkp1) at (0,0) {$A^{(k,k+1)}$};
\node (ikm11) [left=of Akkp1] {$i_{k-1}$};
\node (ikp11) [right=of Akkp1] {$i_{k+1}$};
\node (vk1) [below=of Akkp1,xshift=-1em] {$v_k$};
\node (vkp1) [below=of Akkp1,xshift=1em] {$v_{k+1}$};
\draw (ikm11)--(Akkp1) (ikp11)--(Akkp1) (vk1)--(vk1|-Akkp1.south) (vkp1)--(vkp1|-Akkp1.south);
\end{tikzpicture}\label{merging}\\
\frac{-\partial\mathcal{L}}{\partial A^{(k,k+1)*~w_kw_{k+1}}_{i_{k-1}i_{k+1}}} =
\frac{1}{V}\sum_{(\bm r,  \{\bm n\})\in \mathcal{V}} \frac{\Psi'(\bm r, \{\bm n\}) }{\Psi^{*}(\bm{r}; \{\bm n\})}\nonumber \\ - \frac{\mathcal{N}'}{\mathcal{N}} + \frac{\lambda\tau}{\Tr \left(\rho_{\mathrm{R},k}^2\right)} \label{gradient-main}
\end{gather}
in which 
\begin{align}
\label{eq: Norm}
\mathcal{N}  &= 
\begin{tikzpicture}[baseline=(A1d.north)]
{[start chain]
\node [mat,on chain] (A1) at (0,0){};
\node [on chain, xshift=-0.3em] (cdt1) {$\cdots$};
\node [mat, on chain,xshift=-0.3em] (Ak){};
\node [on chain, xshift=-0.3em] (cdt2) {$\cdots$};
\node [mat, on chain,xshift=-0.3em] (AN){};
}
{[start chain]
\node [mat,on chain, below=of A1] (A1d) {};
\node [on chain, xshift=-0.3em] (cdt1d) {$\cdots$};
\node [mat, on chain,xshift=-0.3em] (Akd) {};
\node [on chain, xshift=-0.3em] (cdt2d) {$\cdots$};
\node [mat, on chain,xshift=-0.3em] (ANd) {};
}
\draw (A1)--(cdt1)--(Ak)--(cdt2)--(AN) (A1d)--(cdt1d)--(Akd)--(cdt2d)--(ANd) (A1)--(A1d) (Ak)--(Akd) (AN)--(ANd);
\end{tikzpicture}
=
\begin{tikzpicture}[baseline=(B.north)]
\node[fmat] (A) at (0,0){$A^{(k)}$};
\draw  (A.east)--++(0.5em,0);
\node[fmat] (B) [below=of A]{$A^{(k)\dagger}$};
\draw  (B.east)--++(0.5em,0);
\draw [] (A)--(B) (A.west)--++(-0.5em,0)|-(B.west) (A.east)--++(0.5em,0)|-(B.east);
\end{tikzpicture}\\
\Tr \left(\rho_{\mathrm{R},k}^2\right) &= 
\begin{tikzpicture}[baseline=(A1p.north)]
{[start chain]
\node[mat, on chain] (A1) at (0,0) {};
\node[on chain, xshift=-0.3em] (cdt1) {$\cdots$};
\node [mat, on chain,xshift=-0.3em] (Ak){};
\node [mat, on chain] (Ak1){};
\node [on chain, xshift=-0.3em] (cdt2) {$\cdots$};
\node [mat, on chain,xshift=-0.3em] (AN){};
}
{[start chain]
\node [mat,on chain, below=of A1] (A1d) {};
\node [on chain, xshift=-0.3em] (cdt1d) {$\cdots$};
\node [mat, on chain,xshift=-0.3em] (Akd) {};
\node [mat, on chain] (Ak1d){};
\node [on chain, xshift=-0.3em] (cdt2d) {$\cdots$};
\node [mat, on chain,xshift=-0.3em] (ANd) {};
}
{[start chain]
\node[mat, on chain, below=of A1d] (A1p) {};
\node[on chain, xshift=-0.3em] (cdt1p) {$\cdots$};
\node [mat, on chain,xshift=-0.3em] (Akp){};
\node [mat, on chain] (Ak1p){};
\node [on chain, xshift=-0.3em] (cdt2p) {$\cdots$};
\node [mat, on chain,xshift=-0.3em] (ANp){};
}
{[start chain]
\node [mat,on chain, below=of A1p] (A1dp) {};
\node [on chain, xshift=-0.3em] (cdt1dp) {$\cdots$};
\node [mat, on chain,xshift=-0.3em] (Akdp) {};
\node [mat, on chain] (Ak1dp){};
\node [on chain, xshift=-0.3em] (cdt2dp) {$\cdots$};
\node [mat, on chain,xshift=-0.3em] (ANdp) {};
}
\draw (A1)--(cdt1)--(Ak)--(Ak1)--(cdt2)--(AN)
(A1d)--(cdt1d)--(Akd)--(Ak1d)--(cdt2d)--(ANd)
(A1p)--(cdt1p)--(Akp)--(Ak1p)--(cdt2p)--(ANp) 
(A1dp)--(cdt1dp)--(Akdp)--(Ak1dp)--(cdt2dp)--(ANdp)
(Ak1.north)--++(0,0.1em)--++(1em,0)|-($(Ak1d.south)-(0,0.1em)$)-|(Ak1d.south)
(Ak1p.north)--++(0,0.1em)--++(1em,0)|-($(Ak1dp.south)-(0,0.1em)$)-|(Ak1dp.south)
(AN.north)--++(0,0.1em)--++(1em,0)|-($(ANd.south)-(0,0.1em)$)-|(ANd.south)
(ANp.north)--++(0,0.1em)--++(1em,0)|-($(ANdp.south)-(0,0.1em)$)-|(ANdp.south)
(A1d)--(A1p)
(Akd)--(Akp)
(A1.north)--++(0,0.1em)--++(-1em,0)|-($(A1dp.south)+(0,-0.1em)$)-|(A1dp.south)
(Ak.north)--++(0,0.1em)--++(-1em,0)|-($(Akdp.south)+(0,-0.1em)$)-|(Akdp.south);
\end{tikzpicture}
= \begin{tikzpicture}[baseline=(Akp.north)]
{[start chain= going below]
\node[fmat, on chain] (Ak) at (0,0) {$A^{(k)\dagger}$};
\node[fmat, on chain] (Akd) {$A^{(k)}$};
\node[fmat, on chain] (Akp) {$A^{(k)\dagger}$};
\node[fmat, on chain] (Akdp) {$A^{(k)}$};
}
\draw (Ak.north)--++(0,0.1em)--++(-2.1em,0)|-($(Akdp.south)-(0,0.1em)$)-|(Akdp.south)
(Ak.west)--++(-0.3em,0)|-(Akdp.west)
(Akd.south)--(Akp.north)
(Akd.west)--++(-0.1em,0)|-(Akp.west)
(Ak.east)--++(0.1em,0)|-(Akd.east)
(Akp.east)--++(0.1em,0)|-(Akdp.east);
\end{tikzpicture}
\end{align}
\begin{align}
&\Psi'(\bm r, \{\bm n\})= \nonumber\\
&\begin{tikzpicture}[baseline=(U1.base)]
{[start chain=1]
\node [fmat,on chain=1] (A1) at (0,0){$A^{(1)\dagger}$};
\node [bguni, above=of A1] (U1) {$U^\dagger_1$};
\node  (v1) [above=of U1]{$r_1$};
\node [on chain=1, xshift=-0.3em] (cdt1) {$\cdots$};
\node [fmat, on chain=1,xshift=-0.3em] (Ak-1){$A^{(k-1)\dagger}$};
}
{[start chain=2]
\node [bguni, on chain, above=0.5em of Ak-1] (Uk-1) {$U^\dagger_{k-1}$};
\node [bguni, on chain] (Uk) {$U^\dagger_{k}$};
\node [bguni, on chain] (Uk+1) {$U^\dagger_{k+1}$};
\node [bguni, on chain] (Uk+2) {$U^\dagger_{k+2}$};
}
\node [above=of Uk-1] (vk-1) [above=0.5em of Uk-1]{$r_{k-1}$};
\node [above=of Uk] (vk){$r_{k}$};
\node [above=of Uk+1] (vk+1){$r_{k+1}$};
\node [above=of Uk+2] (vk+2){$r_{k+2}$};
\draw (Uk.south)--++(0,-0.25em) node[yshift=-0.35em]{$w_k$} (Uk+1.south)--++(0,-0.25em) node[yshift=-0.35em]{$w_{k+1}$};
{[continue chain=1]
\node [fmat, on chain, below=of Uk+2] (Ak+2){$A^{(k+2)\dagger}$};
\node [on chain, xshift=-0.3em] (cdt2) {$\cdots$};
\node [fmat, on chain,xshift=-0.3em] (AN) {$A^{(N)\dagger}$};
\node [bguni, above=of AN] (UN) {$U^\dagger_{N}$};
\node (vN) [above=of UN] {$r_N$};
\draw (Ak-1.east)--++(0.5em, 0) node[right] {$i_{k-1}$}
(Ak+2.west)--++(-0.5em,0) node[left]{$i_{k+1}$};
}
\draw (v1)--(U1)--(A1)
 (vk-1)--(Uk-1)--(Ak-1)
 (vk+2)--(Uk+2)--(Ak+2)
 (vN)--(UN)--(AN) (Uk)--(vk) (Uk+1)--(vk+1);
\draw  (A1)--(cdt1)--(Ak-1) (Ak+2)--(cdt2)--(AN);
\end{tikzpicture}
\end{align}
\begin{align}
\mathcal{N}' &= 
\begin{tikzpicture}[baseline=(A1d.north)]
{[start chain=1]
\node [mat,on chain] (A1) at (0,0){};
\node [mat,on chain] (A2){};
\node [on chain, xshift=-0.3em] (cdt1) {$\cdots$};
\node [mat, on chain,xshift=-0.3em] (Ak-1){};
}
{[start chain=2]
\node [mat,on chain, below=of A1] (A1d) {};
\node [mat,on chain] (A2d){};
\node [on chain, xshift=-0.3em] (cdt1d) {$\cdots$};
\node [mat, on chain,xshift=-0.3em] (Ak-1d){};
}
\node (mg) [mat, minimum width=3.5em, right=of Ak-1]{} ;
{[continue chain=1]
\node [mat, on chain, right=of mg] (Ak+2){};
\node [mat, below=of Ak+2] (Ak+2d){};
\node [on chain, xshift=-0.3em] (cdt2) {$\cdots$};
\node [right=of Ak+2d, xshift=-0.3em] (cdt2d) {$\cdots$};
\node [mat, on chain,xshift=-0.3em] (AN) {};
\node [mat, below=of AN] (ANd) {};
}
\draw (Ak-1d.east)--++(0.5em, 0) node[xshift=0.35em, yshift=-0.35em] {$i_{k-1}$}
(Ak+2d.west)--++(-0.5em,0) node[xshift=-0.35em,yshift=-0.35em]{$i_{k+1}$};
\draw ($(mg.south)+(-0.8em,0)$)--++(0,-0.25em) node[yshift=-0.35em]{$w_k$} ($(mg.south)+(0.8em,0)$)--++(0,-0.25em) node[yshift=-0.35em]{$w_{k+1}$};
\draw (A1)--(A2)--(cdt1)--(Ak-1)--(mg)--(Ak+2)--(cdt2)--(AN)
(A1d)--(A2d)--(cdt1d)--(Ak-1d) (Ak+2d)--(cdt2d)--(ANd)
(A1)--(A1d)
(A2)--(A2d)
(Ak-1)--(Ak-1d)
(Ak+2)--(Ak+2d)
(AN)--(ANd);
\end{tikzpicture}
=
\begin{tikzpicture}[baseline=(mg.base)]
\node (mg) [fmat] at (0,0){$A^{(k,k+1)}$} ;
\draw ($(mg.south)+(-0.8em,0)$)--++(0,-0.25em) node[yshift=-0.35em]{$w_k$}
($(mg.south)+(0.8em,0)$)--++(0,-0.25em) node[yshift=-0.35em]{$w_{k+1}$}
(mg.west)--++(-0.8em,0) node[above]{$i_{k-1}$}
(mg.east)--++(0.8em,0) node[above]{$i_{k+1}$};
\end{tikzpicture}\\
\label{eq: Tprime}
\tau/2 &= 
\begin{tikzpicture}[baseline=(A1p.north)]
{[start chain]
\node[mat, on chain] (A1) at (0,0) {};
\node[on chain, xshift=-0.3em] (cdt1) {$\cdots$};
}
{[start chain]
\node [mat,on chain, below=of A1] (A1d) {};
\node [on chain, xshift=-0.3em] (cdt1d) {$\cdots$};
\node [mat, on chain,xshift=-0.3em, minimum width=3.5em] (mgd) {};
\node [on chain, xshift=-0.3em] (cdt2d) {$\cdots$};
\node [mat, on chain,xshift=-0.3em] (ANd) {};
}
\node [mat, above=of ANd] (AN){};
\node [left=of AN, xshift=0.3em] (cdt2) {$\cdots$};
{[start chain]
\node[mat, on chain, below=of A1d] (A1p) {};
\node[on chain, xshift=-0.3em] (cdt1p) {$\cdots$};
\node [mat, on chain,xshift=-0.3em, minimum width=3.5em] (mgp){};
\node [on chain, xshift=-0.3em] (cdt2p) {$\cdots$};
\node [mat, on chain,xshift=-0.3em] (ANp){};
}
{[start chain]
\node [mat,on chain, below=of A1p] (A1dp) {};
\node [on chain, xshift=-0.3em] (cdt1dp) {$\cdots$};
\node [mat, on chain,xshift=-0.3em, minimum width=3.5em] (mgdp) {};
\node [on chain, xshift=-0.3em] (cdt2dp) {$\cdots$};
\node [mat, on chain,xshift=-0.3em] (ANdp) {};
}
\draw (A1)--(cdt1) (cdt2)--(AN)
(A1d)--(cdt1d)--(mgd)--(cdt2d)--(ANd)
(A1p)--(cdt1p)--(mgp)--(cdt2p)--(ANp) 
(A1dp)--(cdt1dp)--(mgdp)--(cdt2dp)--(ANdp)
($(mgp.north)+(0.8em,0)$)--++(0,0.1em)--++(1em,0)|-($(mgdp.south)+(0.8em,-0.1em)$)-|($(mgdp.south)+(0.8em,0)$)
(ANp.north)--++(0,0.1em)--++(1em,0)|-($(ANdp.south)-(0,0.1em)$)-|(ANdp.south)
(AN.north)--++(0,0.1em)--++(1em,0)|-($(ANd.south)-(0,0.1em)$)-|(ANd.south)
(A1d)--(A1p)
($(mgd.south)-(0.8em,0)$)--($(mgp.north)-(0.8em,0)$)
(A1.north)--++(0,0.1em)--++(-1em,0)|-($(A1dp.south)+(0,-0.1em)$)-|(A1dp.south);
\draw (cdt1.east)--++(0.5em,0) node[above] {$i_{k-1}$}
(cdt2.west)--++(-0.5em,0) node[above] {$i_{k+1}$}
($(mgd.south)+(0.8em,0)$)--++(0,-0.1em)--++(1em,0)--++(0,0.7em) node[above] {$w_{k+1}$}
($(mgdp.south)-(0.8em,0)$)--++(0,-0.1em)--++(-1em,0)--++(0,3.5em) node[above] {$w_{k}$};
\end{tikzpicture}
=
\begin{tikzpicture}[baseline=(Ap.base)]
{[start chain=going below, node distance=1em]
\node[fmat, on chain] (Ad) at (0,0) {$A^{(k,k+1)}$};
\node[fmat, on chain] (Ap) {$A^{(k,k+1)\dagger}$};
\node[fmat, on chain] (Adp) {$A^{(k,k+1)}$};
}
\draw ($(Ad.south)+(-0.8em,0)$)--($(Ap.north)+(-0.8em,0)$)
($(Ad.south)+(0.8em,0)$)--++(0,-0.25em) node[yshift=-0.35em]{$w_{k+1}$}
(Ad.east)--++(0.3em,0) node[above, xshift=0.3em]{$i_{k+1}$}
($(Ap.north)+(0.8em,0)$)--++(0,0.1em)--++(1.3em,0)|-($(Adp.south)+(0.8em,-0.1em)$)-|($(Adp.south)+(0.8em,0)$)
(Ap.east)--++(0.1em,0)|-(Adp.east)
(Ap.west)--++(-0.1em,0)|-(Ad.west)
(Adp.west)--++(-0.3em,0) node[above, xshift=-0.3em]{$i_{k-1}$}
($(Adp.south)-(0.8em,0)$)--++(0,-0.25em) node[yshift=-0.35em]{$w_{k}$};
\end{tikzpicture}
\end{align}

\section{Supplementary Demonstrations for the Assumption about Fidelity Estimation} \label{sec:FidEst}
\begin{figure}[hbtp]
    \centering
    \subfloat[cluster $N=30$]{
    \includegraphics[width=0.3\linewidth]{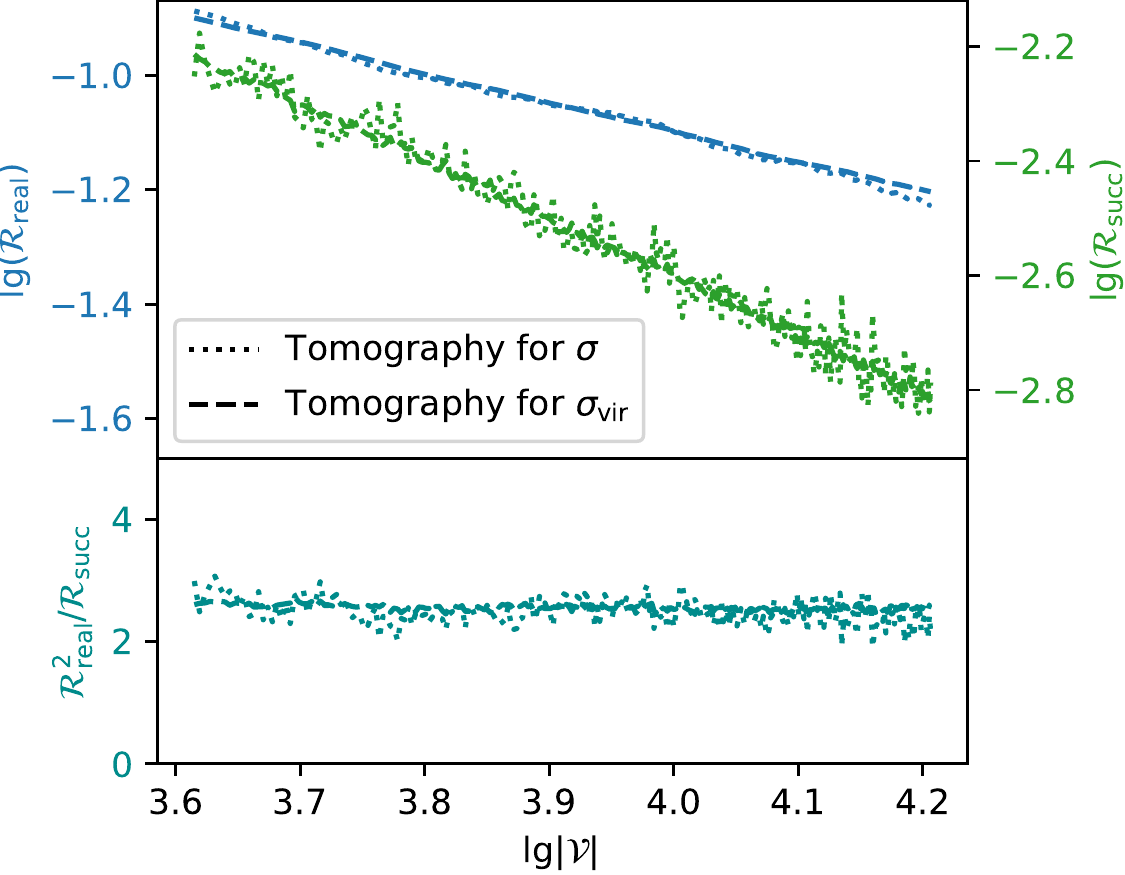}
    }
    \subfloat[dimer $N=30$]{
    \includegraphics[width=0.3\linewidth]{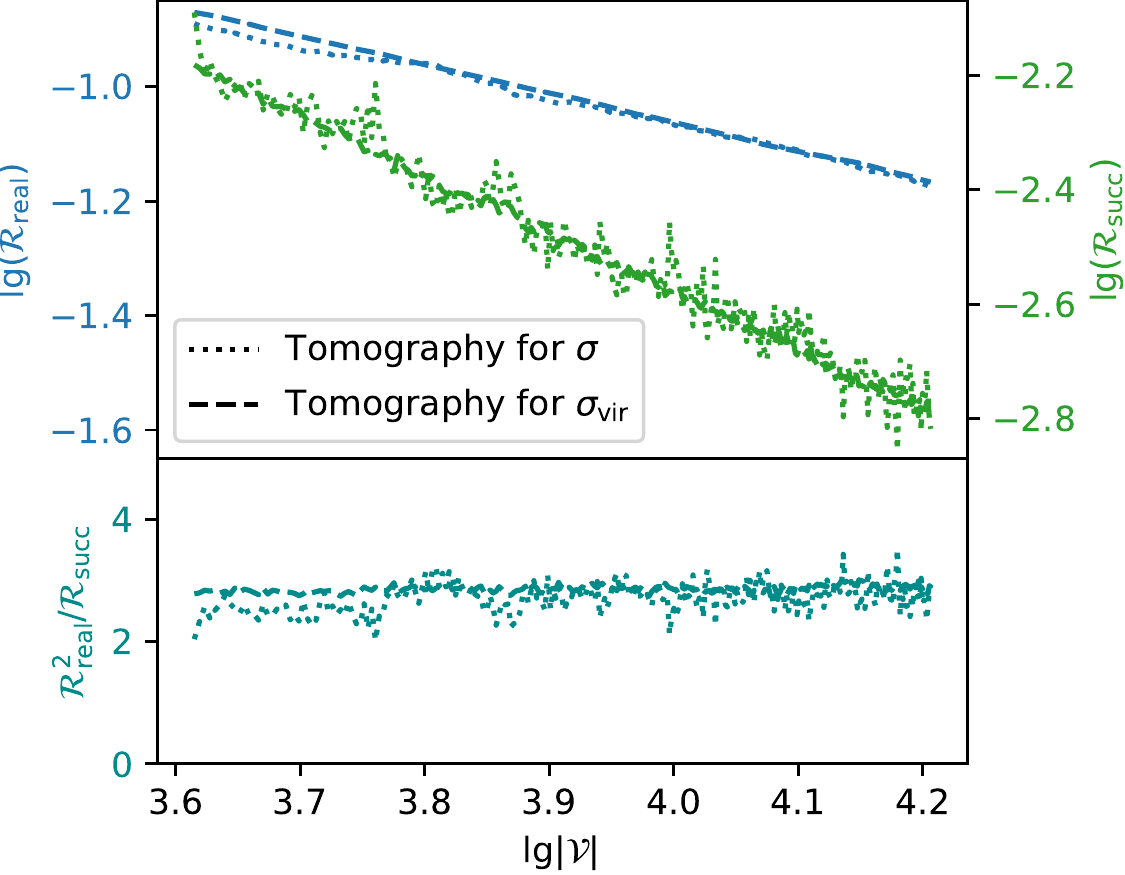}
    }
    \subfloat[random $N=10$ ($D_\mathrm{max}=4$)]{
    \includegraphics[width=0.3\linewidth]{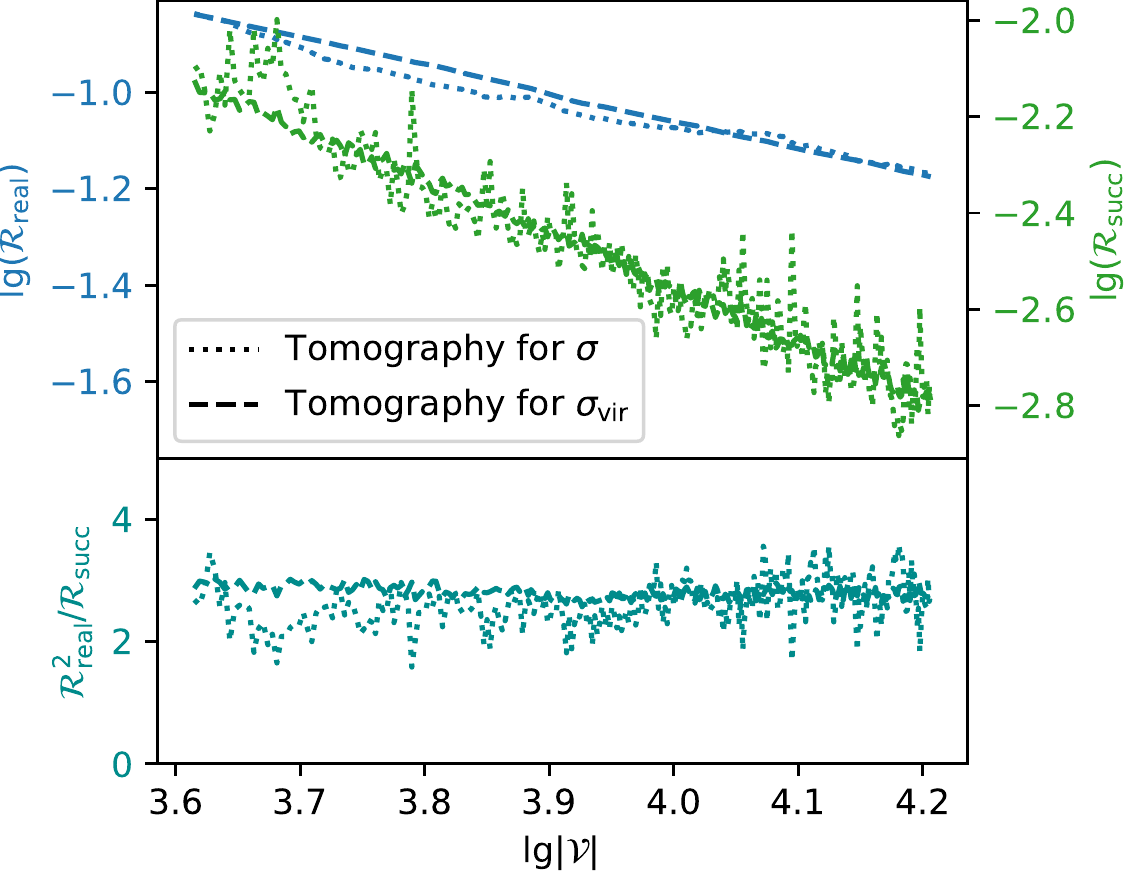}
    }
    \caption{Fidelity estimation in the tomography for three other states. For each one of the three, a MPS is trained up to $|\mathcal{V}|=10000$ outcomes per our scheme, and $\mathcal{R}_\mathrm{real}, \mathcal{R}_\mathrm{succ}$ and $\mathcal{R}_\mathrm{real}^2/\mathcal{R}_\mathrm{succ}$ in this process are plotted in dotted lines, and then we set it as the virtual target state and simulate our scheme in 24 different random cases. We plot the averaged virtual processes in dashed lines.}
    \label{fig:my_label}
\end{figure}

\end{document}